\documentclass[floatfix,twocolumn,superscriptaddress,secnumarabic,amssymb,nobibnotes,amsmath,10pt,aps,prl]{revtex4-1}
\pdfoutput=1
\usepackage{graphicx}
\setkeys{Gin}{width=\columnwidth,keepaspectratio}
\usepackage{transparent}
\usepackage{svgcolor}
\usepackage{calc}
\usepackage[usenames,dvipsnames]{color}
\usepackage{relsize}
\usepackage{import}
\usepackage{ragged2e}
\usepackage{floatrow}
\usepackage{mathtools}
\usepackage{units}
\usepackage{overpic}
\usepackage{printlen}
\uselengthunit{in}
\usepackage{verbatim}
\usepackage{nicefrac}

\usepackage{hyperref}
\usepackage[all]{hypcap}

\renewcommand{\vec}[1]{\ensuremath{\boldsymbol{#1}}}

\renewcommand{\textonehalf}{\nicefrac{1}{2}}

\graphicspath{{./gfx/}}	

\begin{document}
\title{Multi-Q mesoscale magnetism in CeAuSb$_2$}

\author{Guy G. Marcus}
\email[email: ]{guygmarcus@jhu.edu}
\homepage[web: ]{http://www.guygmarcus.com}
\affiliation{Institute for Quantum Matter and Department of Physics and Astronomy,
The Johns Hopkins University, Baltimore, MD 21218, USA}

\author{Dae-Jeong Kim}
\affiliation{Department of Physics and Astronomy, University of California at Irvine, Irvine, California 92697, USA}

\author{Jacob A. Tutmaher}
\affiliation{Institute for Quantum Matter and Department of Physics and Astronomy,
The Johns Hopkins University, Baltimore, MD 21218, USA}

\author{Jose A. Rodriguez-Rivera}
\affiliation{Department of Materials Sciences, University of Maryland, College Park, Maryland 20742, USA}
\affiliation{NIST Center for Neutron Research, Gaithersburg, MD 20899, USA}

\author{Jonas Okkels Birk}
\affiliation{Laboratory for Neutron Scattering and Imaging, Paul Scherrer Institut, CH 5232 Villigen-PSI, Switzerland}
\affiliation{Department of Physics, Technical University of Denmark (DTU), DK-2800 Kgs. Lyngby, Denmark}

\author{Christof Niedermeyer}
\affiliation{Laboratory for Neutron Scattering and Imaging, Paul Scherrer Institut, CH 5232 Villigen-PSI, Switzerland}

\author{Hannoh Lee}
\affiliation{Department of Physics and Astronomy, University of California at Irvine, Irvine, California 92697, USA}

\author{Zachary Fisk}
\affiliation{Department of Physics and Astronomy, University of California at Irvine, Irvine, California 92697, USA}

\author{Collin L. Broholm}
\affiliation{Institute for Quantum Matter and Department of Physics and Astronomy,
The Johns Hopkins University, Baltimore, MD 21218, USA}
\affiliation{NIST Center for Neutron Research, Gaithersburg, MD 20899, USA}

\date{\today}

\begin{abstract}
We report the discovery of a field driven transition from a striped to woven Spin Density Wave (SDW) in the tetragonal heavy fermion compound CeAuSb$_2$. 
Polarized along $\bf c$, the sinusoidal SDW amplitude is 1.8(2)~$\mu_B$/Ce for $T \ll  T_N$=6.25(10)~K with wavevector ${\bf q}_{1}=( \eta, \eta, \textonehalf )$ ($\eta=0.136(2)$). 
For ${\bf H}\parallel{\bf c}$, harmonics appearing at $2{\bf q}_{1}$ evidence a striped magnetic texture below $\mu_\circ H_{c1}=2.78(1)$~T. 
Above $H_{c1}$, these are replaced by woven harmonics at ${\bf q}_{1}+{\bf q}_2=(2\eta, 0, 0)+{\bf c}^*$ until $\mu_\circ H_{c2}=5.42(5)$~T, where satellites vanish and magnetization non-linearly approaches saturation at 1.64(2)~$\mu_B$/Ce for $\mu_\circ H\approx 7$~T.	
\end{abstract}
\pacs{75.25.-j,71.27.+a,72.15.Qm}
\maketitle

From micelles and vesicles in surfactant solutions \cite{Kaler:1989ee, Safran:1990es} to mixed phase type-II superconductors \cite{Blatter:1994gz, Rosenstein:2010bn}, the spontaneous formation of large scale structure in condensed matter is a subject of great beauty, complexity, and practical importance.
The phenomenon is often associated with competing interactions on distinct length scales and sensitivity to external fields that shift a delicate balance.
Heavy fermion systems epitomize this scenario in metals, which place $f$-electrons with strong spin orbit interactions near the transition point between localized and itinerant \cite{Hoshino:2013if, Fisk:1986dj}.
Whether described in terms of oscillatory Ruderman-Kittel-Kasuya-Yosida (RKKY) exchange interactions or Fermi-surface nesting, these strongly interacting Fermi liquids are prone to the development of long wave length modulation of spin, charge, and electronic character with strong sensitivity to applied magnetic fields.

Here we examine the magnetism of the heavy fermion system CeAuSb$_2$, which was previously shown to have two distinct phases versus field ($H$) and temperature ($T$).
By establishing the corresponding magnetic structures, we gain new insight into the interactions and mechanisms that control the phase diagram and give rise to electronic transport anomalies. 
Specifically, we show that the application of a magnetic field along the tetragonal axis of CeAuSb$_2$ induces a transition from a striped to a woven modulation of magnetization, both $\vec{c}$-polarized and modulated on a length scale exceeding the lattice spacing by an order of magnitude. 

\begin{figure}
\centering
	\includegraphics[width=0.95\textwidth]{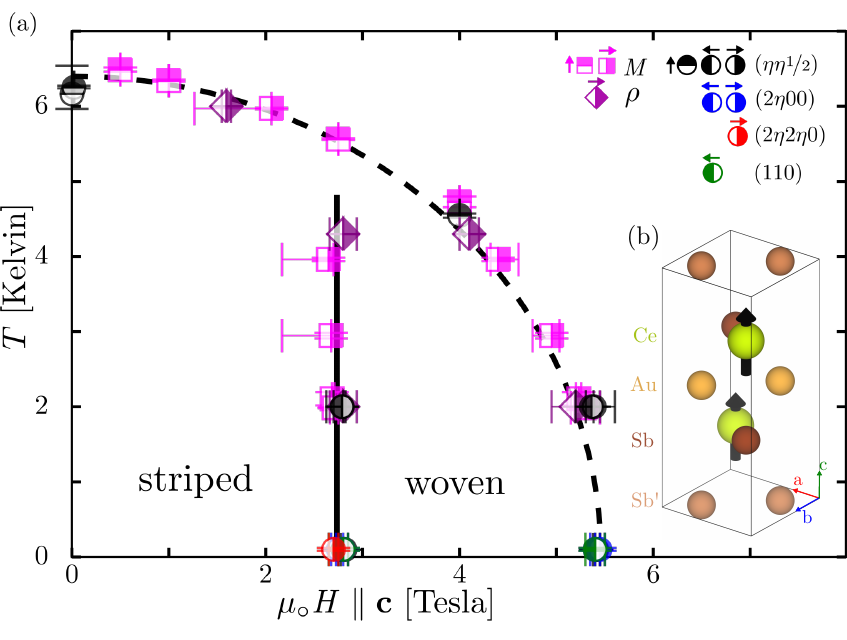}
\caption{
	(a) Phase diagram for CeAuSb$_2$ with boundaries determined from magnetization (squares), resistivity (diamonds) and neutron diffraction (circles). 
	The symbol fill indicates the scan direction within the $H-T$ phase diagram as shown in the legend. 
	(b, inset) Crystallographic unit cell of tetragonal CeAuSb$_2$.
	Magnetic moments shown on Ce sites are associated with the $\Gamma_2$ irreducible representation.
	}
\label{fig:slice_crystal_phase}
\end{figure}

CeAuSb$_2$ is part of the ReTX$_2$ series (Re=La,Ce,Pr,Nd,Sm; T=Cu,Ag,Au; X=Sb,Bi) \cite{Sologub:1994il,Adriano:2014cw,Thomas:2015tya,Seibel:2015go}, which  crystallizes in spacegroup P4/nmm (see \autoref{fig:slice_crystal_phase}(b)). 
Metamagnetic transitions with transport anomalies are common in these compounds, so our findings may have broader relevance. 
CeAuSb$_2$ is Ising-like with an (001) easy axis and lattice parameters $a=4.395$~\AA\ and $c=10.339$~\AA\ at $T=2$~K. 
The Ne\'{e}l temperature is $T_N$=6.25(10)~K and the lower (upper) critical field is $\mu_\circ H_{c1}=2.78(1)$~T ($\mu_\circ H_{c2}=5.42(5)$~T) (\autoref{fig:slice_crystal_phase}(b)) \cite{Balicas:2005eq}.  
We grew high-purity single crystals of CeAuSb$_2$ via the self-flux method and used  8.5(1)~mg and 114.5(1)~mg crystals for diffraction in the ($hh\ell$) and ($hk0$) reciprocal lattice planes, respectively. 
To determine the magnetic structure, we mapped neutron diffraction intensity in the ($hh\ell$) and ($hk0$) planes using the MACS instrument at NIST \cite{Rodriguez:2008be}. 
The sample was rotated by 180 degrees about the vertical axis and the intensity data mapped to one quadrant. 
Field dependence with $\bf H\parallel c$ was studied in the ($hk0$) plane on MACS  with a vertical field magnet and in the ($hh\ell$) plane using RITA-II at PSI \cite{Lefmann:2006ek} using a horizontal field magnet.

\begin{figure}
	\centering
	\includegraphics{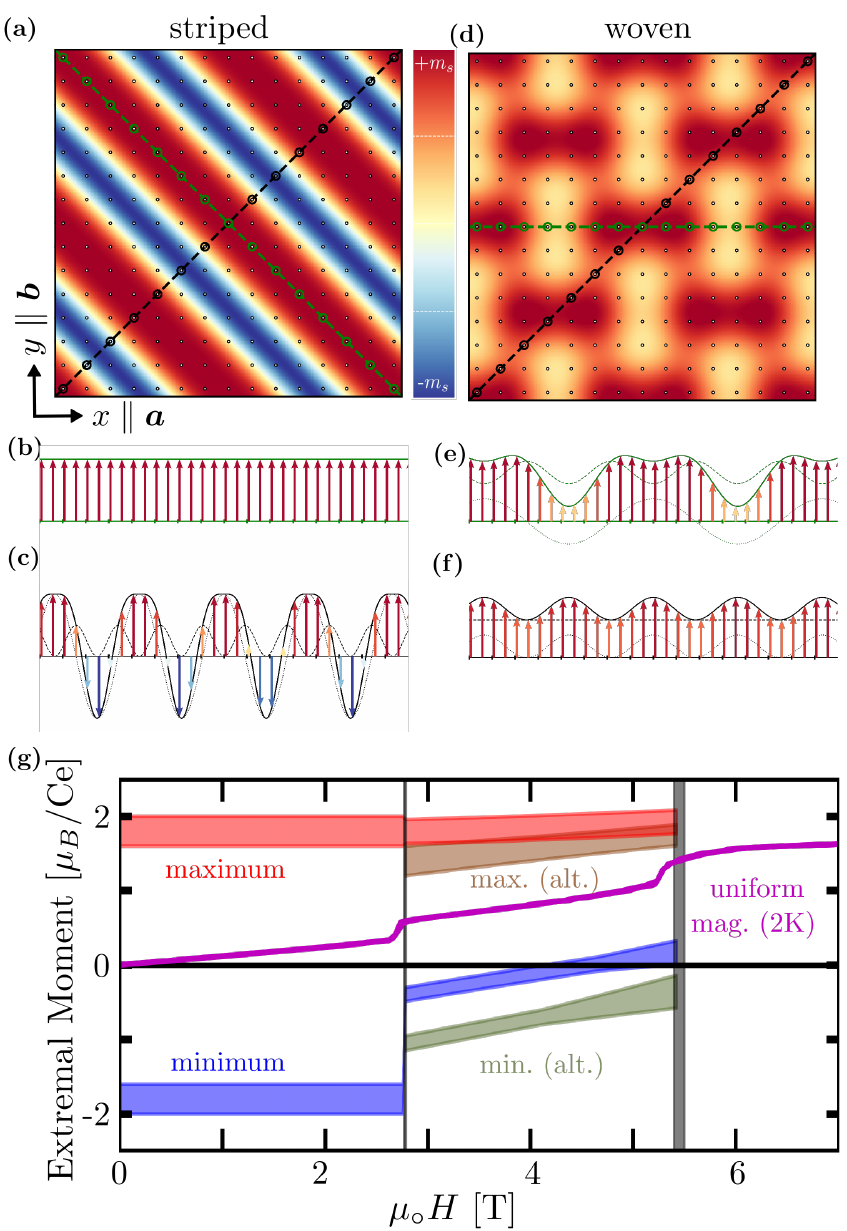}
	\caption{	
	Panels (a-c) show the low field striped magnetic structure and (d-f) show the high field woven structure. 
	Throughout (a-f), the color scale indicates the component of magnetization along $\boldsymbol{c}$ for a single square lattice layer of Ce atoms.
	False color images in (a) and (d) show the magnetic structure within the basal plane while the lower frames (b-c) and (e-f) show the modulation of magnetization along particular lines through the basal plane indicated in frames (a) and (d).
	Panel (g) depicts the maximum and minimum values of local Ce$^{3+}$ magnetization at 100~mK within this model. 
	The maximum values expected from an alternate model (see Supplementary Information) is overlaid for comparison along with the the measured uniform magnetization. 
}
	\label{fig:realspace}
\end{figure}

\begin{figure}
	\centering
	\includegraphics{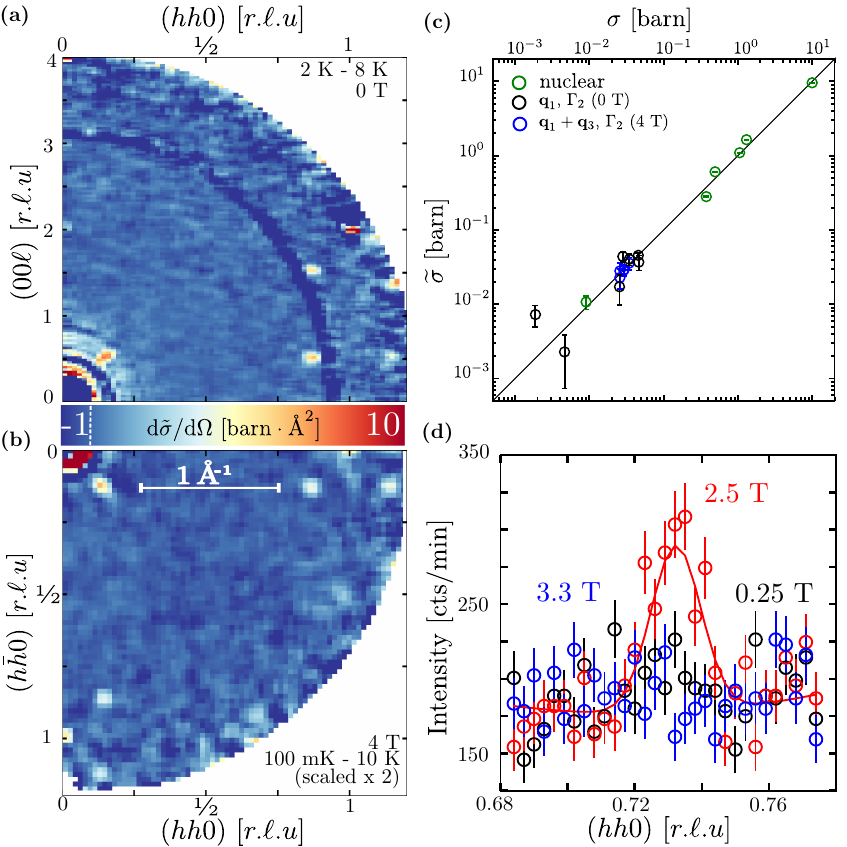}
	\caption{
	Constant field maps of symmetrized, magnetic differential scattering cross-section are shown above for 0~T (a), and 4~T (b). 
	The quality of nuclear and magnetic refinement of these data is demonstrated by a plot of the experimental integrated intensities in absolute units of cross-section ($\widetilde{\sigma}$, see Supplementary Information), versus the calculated cross-section ($\sigma$) in (c).
	Panel (d) shows scans through (110)-$2{\bf q}_1$ at various fields illustrating the appearance of a harmonic peak for intermediate $H$. 
}
	\label{fig:diffraction}
\end{figure}

The difference between diffraction data acquired below (2~K) and above (8~K) $T_N$ is shown in \autoref{fig:diffraction}(a).  
Three out of a quartet of satellite peaks are apparent around (111) and a single satellite is visible near the origin. 
These peaks are indexed by ${\bf q}_1=( \eta  \eta \textonehalf)$ with $\eta=0.136(2)$, indicating a long range ordered magnetic structure that doubles the unit cell along $\bf c$ and is modulated in the basal plane with a wave length $\lambda_m=(a/ \sqrt{2}\eta)=23$~\AA, as depicted in \autoref{fig:realspace}(a). 
The absence of satellite peaks of the form $(\eta,\eta,\textonehalf+n)$, for integer $n \geq 1$ (\autoref{fig:diffraction}(a)) is consistent with diffraction from spins polarized along the $\bf c$-axis. 
To check this hypothesis and establish the size of the ordered moment, we extracted the integrated intensity of the magnetic Bragg peaks by integrating over the relevant areas of the two-dimensional intensity maps. 
The corresponding magnetic diffraction cross sections at $\mu_\circ H$=$0$~T are compared to a striped model with spins oriented along $\bf c$ in \autoref{fig:diffraction}(c), which  provides an excellent account of the  data with a spin density wave amplitude of $m_{{\bf q}_1}=1.8(2)~\mu_B$. 
Here normalization was achieved through comparison to the nuclear diffraction data acquired in the same experiment and compared to expectations for the accepted chemical structure \cite{Sologub:1994il}. 

\begin{figure}[t]
    \label{fig:neutron_thermomag}
    \includegraphics[width=\columnwidth]{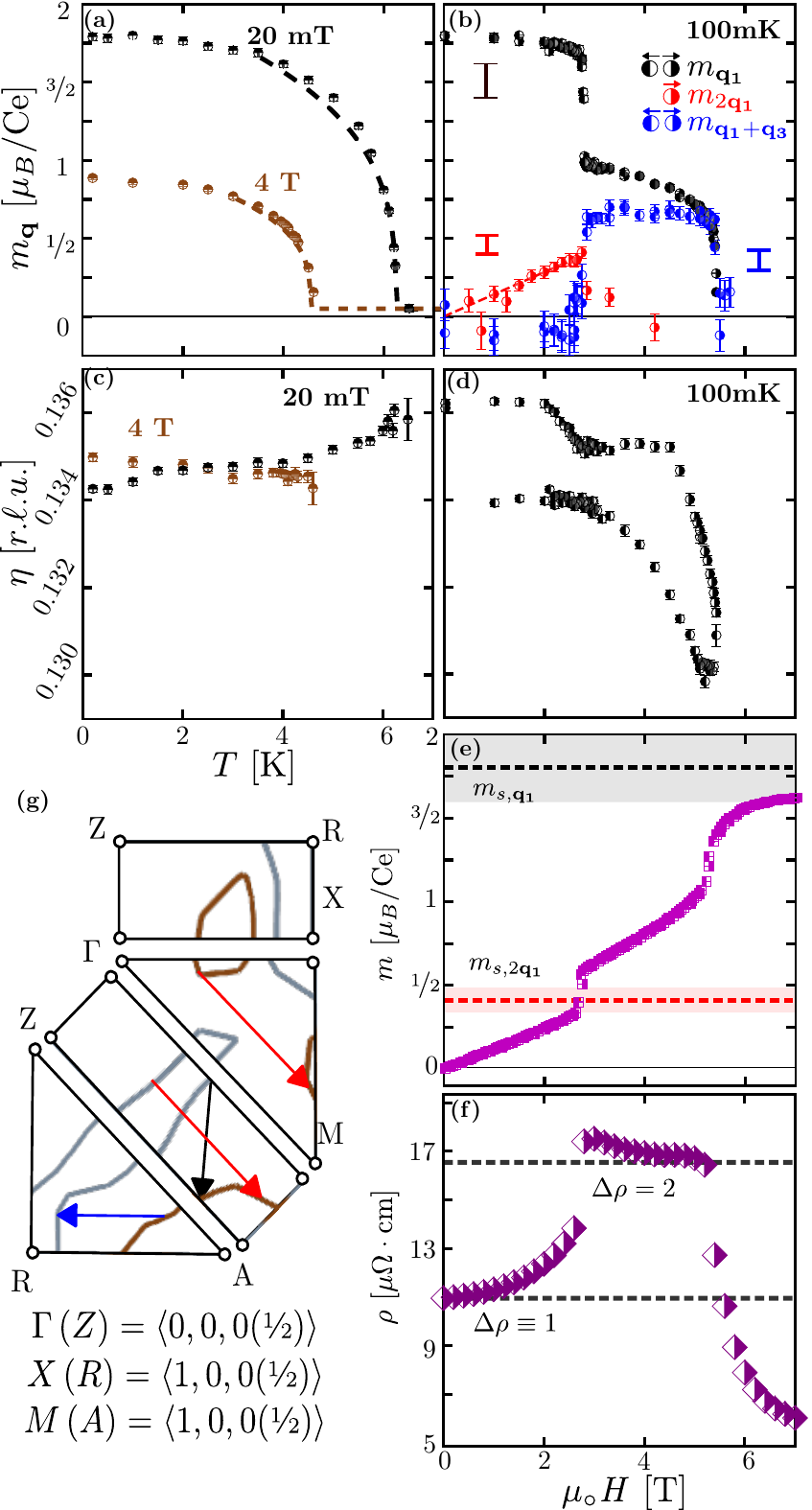}
    \caption{
    Irrep constrained ordered moment (a,b) and ordering wavevector (c,d) is shown here spanning the H-T phase diagram.
    Field dependence of the uniform magnetization is shown in (e).
    Longitudinal magnetoresistance is shown in (f), with dashed lines highlighting a factor two increase of $\Delta\rho$ across $\mathit{H}_{c1}$.
	The reduced Fermi surface (g) is extracted from DFT calculations and overlaid with potential nesting conditions for {\bf q}$_1$ (black), 2{\bf q}$_1$ (red), and ${\bf q}_1+ {\bf q}_2$ (blue).
    }
\end{figure}

\autoref{fig:neutron_thermomag}(a) reports the ordered moment versus $T$ as extracted from the wave vector integrated magnetic neutron diffraction intensity at  $( 1-\eta,1-\eta,\textonehalf )$ and for $\mu_\circ H$=20~mT.
Near $T_N$ these data can be described as $m_{{\bf q}_1}(T)\propto(1-T/T_N)^\beta$ where $\beta$=0.32(5), consistent with the  $\beta$=0.326 for the 3D Ising model \cite{Pelissetto2002549}, but also with $\beta=0.3645(25)$ for the 3D Heisenberg model \cite{PhysRevB.21.3976}.
Landau theory at such a second order phase transition predicts the magnetic structure forms a single irreducible representation (IR) of the little group, $\bf{G}_{\bf q}$, associated with ${\bf q}_1$. 
A description of the diffraction data in \autoref{fig:diffraction}(a) by either $\Gamma_2$ ($\uparrow\uparrow$) or $\Gamma_3$ ($\uparrow\downarrow$) is consistent with this tenet. 
Application of the P4/nmm symmetry operations  generates a second distinct wavevector (arm) of the star \{${\bf q}_i$\} namely ${\bf {q}}_2=(   \eta \bar{\eta} \textonehalf)$. 
The observation of ${\bf q}_1$ satellite peaks leaves open whether distinct, single-{\bf q} domains or a multi-{\bf q} modulation describes the zero-field magnetic structure. 
As we shall now show, this is resolved by analysis of magnetic diffraction data in a field ${\bf H}\parallel{\bf c}$. 

We enter the striped phase by zero-field cooling (ZFC) to 100~mK.
Initial application of a magnetic field along $\bf{c}$ has little effect on the ordered moment $m_{{\bf q}_1}$ until an abrupt reduction of 0.65(5)$\mu_B$ at $\mu_\circ H_{c1}$=2.78(1)~T (\autoref{fig:neutron_thermomag}(b)).
Continuing this isothermal field-sweep (IFS) to higher fields, the staggered magnetization $m_{{\bf q}_1}$is continually suppressed as in the approach to a second order phase transition before eventually falling below the detection limit above  $\mu_\circ H_{c2}$=5.42(5)~T.

While no hysteresis was detected in the field dependence of the ordered moment, we do find hysteresis in the field dependence of the characteristic wavevector.
\autoref{fig:neutron_thermomag}(d) shows that $\eta$ locks into two distinct plateaus for increasing IFS each terminated by regimes where $\eta$, to within resolution, decreases continuously with increasing $H$. 
For decreasing IFS, $\eta$ follows a different, non-intersecting trajectory without plateaus. 
This hysteresis in ${\bf q}_1$, and in the higher harmonics discussed below,  persists to the lowest fields and for $T$s up to at least 2~K (see Supplementary Information), while no hysteresis is observed in the field dependence of the staggered magnetization nor in the uniform magnetization, $m_\circ$ (\autoref{fig:neutron_thermomag}(e)). 

$m_\circ$ increases linearly with applied field at a rate of $m_\circ^\prime=0.12(1)\mu_B {\rm T}^{-1}/$Ce until an abrupt increase of $\Delta m_\circ=0.23(3)\mu_B$/Ce at $H_{c1}$.
Above this transition, $m_\circ$ continues to increase linearly at a similar rate until $H_{c2}$, where the incommensurate magnetic peaks vanish. 
Interestingly,  $m_\circ$ continues to increase for $H>H_{c2}$ until saturating at $1.64(1)~\mu_B$/Ce, which is indistinguishable from the zero field staggered magnetization (gray band in \autoref{fig:neutron_thermomag}(e)).   

\autoref{fig:neutron_thermomag}(f) shows the longitudinal magnetoresistivity versus $H$ at 2~K where the $\rho(T)$ is dominated by the residual component. 
For $H=H_{c1}$ there is an abrupt increase in resistivity that is subsequently reversed for $H>H_{c2}$.
This decrease in $\rho$ is approximately twice as large as the increase in $\rho$ at $H_{c1}$. 
One interpretation is that parts of the Fermi surface develop a gap in the ordered regimes. 

\autoref{fig:diffraction}(d) shows representative line cuts of elastic neutron scattering along ($hh0$) for $ H<H_{c1}$. 
We find a weak, field-induced peak at $( 2\eta,2\eta,0)=2{\bf q}_1-{\bf c}^*$, which indicates the spatial modulation of magnetization ceases to follow a simple sinusoidal form in a field. The new Fourier component is supported by a single ${\bf q}_1$ domain and is not accompanied by harmonics of the form  ${\bf q}_1\pm {\bf q}_2$. 
This constitutes evidence that the $H < H_{c1}$ SDW state is striped and consists of distinct ${\bf q}_1$ and ${\bf q}_2$ domains. Furthermore, the presence of a magnetic satellite peak at momentum transfer ${\bf Q}=\left(2\eta,2\eta,0 \right)$ implies the SDW moment associated with the two Ce$^{3+}$ spins within a unit cell are in phase corresponding to the $\Gamma_2$ ($\uparrow\uparrow$) IR.

The field dependence of the amplitude of the $2{\bf q}_1$ harmonic, $m_{2{\bf q}_1}(H)$, is shown in \autoref{fig:neutron_thermomag}(b). 
As directly apparent from \autoref{fig:diffraction}, there is no evidence for this harmonic in zero field with a quantitative limit of $|m_{3{\bf q}_1}|<0.1~\mu_B$. 
Upon comparison to the 0.60(7)$\mu_B$ third order harmonic expected from a square-wave structure, this indicates a sinusoidal modulation for $H=0$. 
A linear in $H$ fit to $m_{2{\bf q}_1}(H)$ yields $m^\prime_{2{\bf q}_1}=0.14(2)\mu_B {\rm T}^{-1}/$Ce, which is indistinguishable from $m_\circ^\prime$ so that  $|m_{2{\bf q}_1}(H)|\approx |m_0(H)|$ throughout the striped phase  (\autoref{fig:neutron_thermomag}(b,e)). 
Combining the three Fourier components we obtain $m_j({\bf r})=m_0+\nu^j m_{{\bf q}_1}\cos({\bf q}_1\cdot{\bf r})+m_{2{\bf q}_1}\cos (2{\bf q}_1\cdot{\bf r})$ on sublattice $j$. Here $\nu=1$ (IR: $\Gamma_2$) and $\nu=-1$ (IR: $\Gamma_3$) cannot be distinguished in the present data. 

Without loss of generality we pick $m_0>0$ and $m_{{\bf q}_1}>0$. To ensure $|m_j({\bf r})|$ does not exceed the saturation magnetization at any ${\bf r}$ requires $m_{2{\bf q}_1}<0$. The corresponding $m_j({\bf r})=\nu^j m_{{\bf q}_1}\cos({\bf q}_1\cdot{\bf r})+m_\circ(1-\cos (2{\bf q}_1\cdot{\bf r}))$ for $H$ immediately below $H_{c1}$ is shown in \autoref{fig:slice_crystal_phase}(c-e). Here we have used our experimental finding that $m_0(H)\approx-m_{2{\bf q}_1}(H)$.
Qualitatively, we find stripes where $m({\bf r})>0$  have broadened at the expense of stripes where $m_j({\bf r})<0$.
Given only the fundamental and the first harmonics and assuming $m_{2{\bf q}_1}=-m_0$, a global maximum in $m_j({\bf r})$ that exceeds $m_{{\bf q}_1}$ occurs when $m_0$ exceeds $m_{{\bf q}_1}/4$. 
The similarity of $m_0(H_{c1})=0.32~\mu_B$ to $m_{{\bf q}_1}(H_{c1})/4=0.43(1)~\mu_B$ indicates the phase transition at $H_{c1}$ is  associated with reaching the maximum magnetization possible for a striped phase dominated by just three Fourier components $m_0$, $m_{{\bf q}_1}$, and $m_{2{\bf q}_1}$.

\autoref{fig:neutron_thermomag}(b) shows $m_{2{\bf q}_1}$ abruptly vanishes for $H>H_{c1}$. 
The false color map of the ($hk0$) plane at $\mu_0H=4$~T in \autoref{fig:diffraction}(b) shows the $2{\bf q}_1$ harmonic is replaced by satellites spanned by ${\bf q}_1\pm{\bf q}_2$ that surround  ($110$), ($1\bar{1}0$), and $(000)$. 
These indicate the simultaneous presence at the atomic scale of both ${\bf q}_1$ and ${\bf q}_2$ and a field induced harmonic that transforms as $\Gamma_2$. 
\autoref{fig:neutron_thermomag} shows $m_{{\bf q}_1\pm {\bf q}_2}$, abruptly jumps to and then holds an essentially constant value of 0.7(1)$\mu_B$/Ce for $H_{c1}<H<H_{c2}$. 
The similarity to the plateau-like dependence of the residual magneto-resistivity is consistent with both phenomena arising from the opening of an additional gap on the Fermi surface: Two nesting wavevectors, rather than one, gap out twice as much of the Fermi surface, thereby doubling the residual resistivity as observed (\autoref{fig:neutron_thermomag}(f)). 

For $H_{c1}<H<H_{c2}$ the ${\bf c}$-oriented staggered magnetization can be described as 
\begin{eqnarray}
m_j({\bf r})&=&m_0+\tfrac{1}{2} \nu^j m_{{\bf q}_{1}}(\cos({\bf q}_1\cdot{\bf r})+\cos({\bf q}_2\cdot{\bf r}))\\
&+&\tfrac{1}{2}m_{{\bf q}_{1}\pm{\bf q}_{2}}(\cos(({\bf q}_1+{\bf q}_2)\cdot{\bf r})+\delta \cos(({\bf q}_1-{\bf q}_2)\cdot{\bf r})).\nonumber
\label{magform}
\end{eqnarray}
Here $m_0$ and $m_{{\bf q}_{1}}=m_{{\bf q}_{2}}$ can again be chosen positive without loss of generality. 
There are two qualitatively different structures $\delta=\pm 1$ that we distinguish by considering the field dependent amplitudes shown in \autoref{fig:neutron_thermomag}(b). 
For $\delta=1$, $m({\bf r})$ describes a checkered pattern with four fold symmetry and $m_{{\bf q}_{1}\pm{\bf q}_{2}}<0$ to avoid exceeding the saturation moment on any site (see Supplementary Information). 
However for this structure, the maximum magnetization of any Ce$^{3+}$ site falls considerably below saturation immediately above $H_{c1}$ (\autoref{fig:realspace}(g)). 
While this might be possible if gradient terms dominate over quartic terms in a Landau free energy, the choice of $\delta=-1$, labeled as the ``woven'' phase, has the virtue that max$[m_j({\bf r})]$ remains virtually constant through $H_{c1}$ and all the way up to $H_{c2}$ when the measured field dependent amplitudes and magnetization in \autoref{fig:neutron_thermomag}(b,e) are considered in \autoref{magform}.
While diffraction cannot provide definite proof for this structure, illustrated in \autoref{fig:realspace}(d-g), there is circumstantial evidence. 

\autoref{fig:realspace}(d) and Eq.~\ref{magform} show that the woven SDW, just as the crystal structure, is not invariant under $C_4$: Lobes of $\vec{c}$-polarized spins extend along ${\bf a}$ (${\bf b}$) for $m_{{\bf q}_{1}\pm{\bf q}_{2}}>0$ ($m_{{\bf q}_{1}\pm{\bf q}_{2}}<0$). 
Either between sublattices or at the transition between unit cells along $\bf c$, the woven pattern shifts within the basal plane by half of its period in the direction of the prolate axis of lobes with magnetization antiparallel to $\vec{H}$.
As was the case for $H<H_{c1}$, there are \emph{two} spatially separated domains only now composed of $\vec{c}$-polarized lobes of spins extended along $\bf a$ or $\bf b$.  
However, the development of magnetization in the woven structure is qualitatively distinct. 
This is apparent in \autoref{fig:neutron_thermomag} where the fundamental amplitude $m_{{\bf q}_1}$ decreases with field at a rate of $m^\prime_{{\bf q}_1}=-0.13(1)\mu_B$~T$^{-1}/$~Ce while the harmonic $m_{{\bf q}_1\pm{\bf q}_2}$ is field independent.   
$m_\circ^\prime=0.12(1)\mu_B$~T$^{-1}/$~Ce maintains the same value in the woven phase as it had in the striped phase. 
\autoref{fig:realspace}(g) shows this corresponds to increasing the magnetization of negatively magnetized regions only. 

Throughout the magnetization process shown in \autoref{fig:neutron_thermomag}(c-d), the magnetic wavelength, $\lambda_m$, shows less than a 5\% variation. 
This contrasts with other cerium-based Ising systems.
For example, CeSb undergoes a series of field driven phase transitions that alter the direction of magnetization of entire planes of spins from  $\uparrow\uparrow\downarrow\downarrow$ ($q=(001/2)$), to $\uparrow\uparrow\downarrow\downarrow\uparrow\uparrow\downarrow$ ($q=(004/7)$), to $\uparrow\uparrow\downarrow\uparrow\uparrow\downarrow$ ($q=(002/3)$), all the way to full ferromagnetism ($q=0$) \cite{PhysRevB.49.15096}. 
These square-wave structures are modulated along the easy-axis and can be accounted for by the ANNNI (Anisotropic Nearest and Next Nearest Neighbor Ising) model, where 4f electrons are described as localized Ising degrees of freedom subject to oscillatory, anisotropic RKKY interactions. 

A model of competing near neighbor exchange interactions that reproduces the critical wave vectors ${\bf q}_1$ and ${\bf q}_2$ as well as the Weiss temperature and the upper critical field is possible. 
At a minimum, it involves antiferromagnetic $J_c$>0 between Ce sites on the $\vec{c}$-bond and basal-plane interactions $J_1>0$ on the $\vec{a}$-bond, $J_2<J_1/4$ on the ($\vec{a}+\vec{b}$)-bond, and $J_3=-J_1/4\cos2\pi\eta$ on the 2$\vec{a}$-bond \cite{Seabra:2016ba}. 
However, the absence of harmonics at zero field and low temperatures appears inconsistent with such a framework and points to a Fermi surface nesting induced SDW. 
To examine this possibility, we calculated the Fermi surface using the generalized gradient approximation (see Supplementary Information).
The result is shown in \autoref{fig:neutron_thermomag}(g). 
Near the Fermi level, the band structure is dominated by $f$-electrons with contributions to the low energy density of states from sharply dispersing $p$-bands an order of magnitude smaller.
While there are no ideal nesting conditions, ${\bf q}_1$, $2{\bf q}_1$, and ${\bf q}_1\pm{\bf q}_2$ do connect areas of the $f$-electron dominated Fermi-surface, consistent with an SDW instability. 

The distinct hysteresis of the small changes in SDW wave vector versus field (\autoref{fig:neutron_thermomag}(c,d)) is indicative of the profound rearrangement of static magnetism at $H_{c1}$ and $H_{c2}$. Upon reducing field at low $T$, nucleation of the woven state from the paramagnetic state at $H_{c2}$ can be expected to allow for greater adherence to constraints imposed by impurities and defects than when nucleating the woven state within the striped state upon increasing field past $H_{c1}$ at low $T$. Within the SDW picture the corresponding subtle differences in magnetic order provide a natural explanation for field-hysteretic electronic transport.

Our results provide a simple phenomenological description of the magnetization process in CeAuSb$_2$ that determines the critical magnetization at the metamagnetic transitions. Net magnetization is achieved by adding both a uniform and a single first harmonic component to a sinusoidal magnetization wave while maintaining the fundamental wave length and maximum amplitude. 
A SDW picture would appear to be appropriate and might allow for a unified understanding of the many meta-magnetic transitions in the ReTX$_2$ family of heavy fermion compounds. 

We are glad to thank Christian Batista, Martin Mourigal, Sid Parameswaran, Chandra Varma, Yuan Wan, and Andrew Wills for helpful discussions. This research was funded by the U.S. Department of Energy, Office of Basic Science, Division of Materials Sciences and Engineering, Grant No. DE-FG02-08ER46544. GGM acknowledges generous support from the NSF-GRFP, Grant No. DGE-1232825. 

\bibliography{ceausb2_a}

\begin{thebibliography}{17}%
\makeatletter
\providecommand \@ifxundefined [1]{%
 \@ifx{#1\undefined}
}%
\providecommand \@ifnum [1]{%
 \ifnum #1\expandafter \@firstoftwo
 \else \expandafter \@secondoftwo
 \fi
}%
\providecommand \@ifx [1]{%
 \ifx #1\expandafter \@firstoftwo
 \else \expandafter \@secondoftwo
 \fi
}%
\providecommand \natexlab [1]{#1}%
\providecommand \enquote  [1]{``#1''}%
\providecommand \bibnamefont  [1]{#1}%
\providecommand \bibfnamefont [1]{#1}%
\providecommand \citenamefont [1]{#1}%
\providecommand \href@noop [0]{\@secondoftwo}%
\providecommand \href [0]{\begingroup \@sanitize@url \@href}%
\providecommand \@href[1]{\@@startlink{#1}\@@href}%
\providecommand \@@href[1]{\endgroup#1\@@endlink}%
\providecommand \@sanitize@url [0]{\catcode `\\12\catcode `\$12\catcode
  `\&12\catcode `\#12\catcode `\^12\catcode `\_12\catcode `\%12\relax}%
\providecommand \@@startlink[1]{}%
\providecommand \@@endlink[0]{}%
\providecommand \url  [0]{\begingroup\@sanitize@url \@url }%
\providecommand \@url [1]{\endgroup\@href {#1}{\urlprefix }}%
\providecommand \urlprefix  [0]{URL }%
\providecommand \Eprint [0]{\href }%
\providecommand \doibase [0]{http://dx.doi.org/}%
\providecommand \selectlanguage [0]{\@gobble}%
\providecommand \bibinfo  [0]{\@secondoftwo}%
\providecommand \bibfield  [0]{\@secondoftwo}%
\providecommand \translation [1]{[#1]}%
\providecommand \BibitemOpen [0]{}%
\providecommand \bibitemStop [0]{}%
\providecommand \bibitemNoStop [0]{.\EOS\space}%
\providecommand \EOS [0]{\spacefactor3000\relax}%
\providecommand \BibitemShut  [1]{\csname bibitem#1\endcsname}%
\let\auto@bib@innerbib\@empty
\bibitem [{\citenamefont {Kaler}\ \emph {et~al.}(1989)\citenamefont {Kaler},
  \citenamefont {Murthy}, \citenamefont {Rodriguez},\ and\ \citenamefont
  {Zasadzinski}}]{Kaler:1989ee}%
  \BibitemOpen
  \bibfield  {author} {\bibinfo {author} {\bibfnamefont {E.~W.}\ \bibnamefont
  {Kaler}}, \bibinfo {author} {\bibfnamefont {A.~K.}\ \bibnamefont {Murthy}},
  \bibinfo {author} {\bibfnamefont {B.~E.}\ \bibnamefont {Rodriguez}}, \ and\
  \bibinfo {author} {\bibfnamefont {J.~A.}\ \bibnamefont {Zasadzinski}},\
  }\href@noop {} {\bibfield  {journal} {\bibinfo  {journal} {Science}\ }\textbf
  {\bibinfo {volume} {245}},\ \bibinfo {pages} {1371} (\bibinfo {year}
  {1989})}\BibitemShut {NoStop}%
\bibitem [{\citenamefont {Safran}\ \emph {et~al.}(1990)\citenamefont {Safran},
  \citenamefont {Pincus},\ and\ \citenamefont {Andelman}}]{Safran:1990es}%
  \BibitemOpen
  \bibfield  {author} {\bibinfo {author} {\bibfnamefont {S.~A.}\ \bibnamefont
  {Safran}}, \bibinfo {author} {\bibfnamefont {P.}~\bibnamefont {Pincus}}, \
  and\ \bibinfo {author} {\bibfnamefont {D.}~\bibnamefont {Andelman}},\
  }\href@noop {} {\bibfield  {journal} {\bibinfo  {journal} {Science}\ }\textbf
  {\bibinfo {volume} {248}},\ \bibinfo {pages} {354} (\bibinfo {year}
  {1990})}\BibitemShut {NoStop}%
\bibitem [{\citenamefont {Blatter}\ \emph {et~al.}(1994)\citenamefont
  {Blatter}, \citenamefont {Feigel'man}, \citenamefont {Geshkenbein},
  \citenamefont {Larkin},\ and\ \citenamefont {Vinokur}}]{Blatter:1994gz}%
  \BibitemOpen
  \bibfield  {author} {\bibinfo {author} {\bibfnamefont {G.}~\bibnamefont
  {Blatter}}, \bibinfo {author} {\bibfnamefont {M.~V.}\ \bibnamefont
  {Feigel'man}}, \bibinfo {author} {\bibfnamefont {V.~B.}\ \bibnamefont
  {Geshkenbein}}, \bibinfo {author} {\bibfnamefont {A.~I.}\ \bibnamefont
  {Larkin}}, \ and\ \bibinfo {author} {\bibfnamefont {V.~M.}\ \bibnamefont
  {Vinokur}},\ }\href@noop {} {\bibfield  {journal} {\bibinfo  {journal}
  {Reviews of Modern Physics}\ }\textbf {\bibinfo {volume} {66}},\ \bibinfo
  {pages} {1125} (\bibinfo {year} {1994})}\BibitemShut {NoStop}%
\bibitem [{\citenamefont {Rosenstein}\ and\ \citenamefont
  {Li}(2010)}]{Rosenstein:2010bn}%
  \BibitemOpen
  \bibfield  {author} {\bibinfo {author} {\bibfnamefont {B.}~\bibnamefont
  {Rosenstein}}\ and\ \bibinfo {author} {\bibfnamefont {D.}~\bibnamefont
  {Li}},\ }\href@noop {} {\bibfield  {journal} {\bibinfo  {journal} {Reviews of
  Modern Physics}\ }\textbf {\bibinfo {volume} {82}},\ \bibinfo {pages} {109}
  (\bibinfo {year} {2010})}\BibitemShut {NoStop}%
\bibitem [{\citenamefont {Hoshino}\ and\ \citenamefont
  {Kuramoto}(2013)}]{Hoshino:2013if}%
  \BibitemOpen
  \bibfield  {author} {\bibinfo {author} {\bibfnamefont {S.}~\bibnamefont
  {Hoshino}}\ and\ \bibinfo {author} {\bibfnamefont {Y.}~\bibnamefont
  {Kuramoto}},\ }\href@noop {} {\bibfield  {journal} {\bibinfo  {journal}
  {Physical Review Letters}\ }\textbf {\bibinfo {volume} {111}},\ \bibinfo
  {pages} {026401} (\bibinfo {year} {2013})}\BibitemShut {NoStop}%
\bibitem [{\citenamefont {Fisk}\ \emph {et~al.}(1986)\citenamefont {Fisk},
  \citenamefont {Ott}, \citenamefont {Rice},\ and\ \citenamefont
  {Smith}}]{Fisk:1986dj}%
  \BibitemOpen
  \bibfield  {author} {\bibinfo {author} {\bibfnamefont {Z.}~\bibnamefont
  {Fisk}}, \bibinfo {author} {\bibfnamefont {H.}~\bibnamefont {Ott}}, \bibinfo
  {author} {\bibfnamefont {T.}~\bibnamefont {Rice}}, \ and\ \bibinfo {author}
  {\bibfnamefont {J.}~\bibnamefont {Smith}},\ }\href@noop {} {\bibfield
  {journal} {\bibinfo  {journal} {Nature}\ }\textbf {\bibinfo {volume} {320}},\
  \bibinfo {pages} {124} (\bibinfo {year} {1986})}\BibitemShut {NoStop}%
\bibitem [{\citenamefont {Sologub}\ \emph {et~al.}(1994)\citenamefont
  {Sologub}, \citenamefont {Hiebl}, \citenamefont {Rogl}, \citenamefont
  {No{\"e}l},\ and\ \citenamefont {Bodak}}]{Sologub:1994il}%
  \BibitemOpen
  \bibfield  {author} {\bibinfo {author} {\bibfnamefont {O.}~\bibnamefont
  {Sologub}}, \bibinfo {author} {\bibfnamefont {K.}~\bibnamefont {Hiebl}},
  \bibinfo {author} {\bibfnamefont {P.}~\bibnamefont {Rogl}}, \bibinfo {author}
  {\bibfnamefont {H.}~\bibnamefont {No{\"e}l}}, \ and\ \bibinfo {author}
  {\bibfnamefont {O.}~\bibnamefont {Bodak}},\ }\href@noop {} {\bibfield
  {journal} {\bibinfo  {journal} {Journal of Alloys and Compounds}\ }\textbf
  {\bibinfo {volume} {210}},\ \bibinfo {pages} {153} (\bibinfo {year}
  {1994})}\BibitemShut {NoStop}%
\bibitem [{\citenamefont {Adriano}\ \emph {et~al.}(2014)\citenamefont
  {Adriano}, \citenamefont {Rosa}, \citenamefont {Jesus}, \citenamefont
  {Mardegan}, \citenamefont {Garitezi}, \citenamefont {Grant}, \citenamefont
  {Fisk}, \citenamefont {Garcia}, \citenamefont {Reyes}, \citenamefont {Kuhns},
  \citenamefont {Urbano}, \citenamefont {Giles},\ and\ \citenamefont
  {Pagliuso}}]{Adriano:2014cw}%
  \BibitemOpen
  \bibfield  {author} {\bibinfo {author} {\bibfnamefont {C.}~\bibnamefont
  {Adriano}}, \bibinfo {author} {\bibfnamefont {P.~F.~S.}\ \bibnamefont
  {Rosa}}, \bibinfo {author} {\bibfnamefont {C.~B.~R.}\ \bibnamefont {Jesus}},
  \bibinfo {author} {\bibfnamefont {J.~R.~L.}\ \bibnamefont {Mardegan}},
  \bibinfo {author} {\bibfnamefont {T.~M.}\ \bibnamefont {Garitezi}}, \bibinfo
  {author} {\bibfnamefont {T.}~\bibnamefont {Grant}}, \bibinfo {author}
  {\bibfnamefont {Z.}~\bibnamefont {Fisk}}, \bibinfo {author} {\bibfnamefont
  {D.~J.}\ \bibnamefont {Garcia}}, \bibinfo {author} {\bibfnamefont {A.~P.}\
  \bibnamefont {Reyes}}, \bibinfo {author} {\bibfnamefont {P.~L.}\ \bibnamefont
  {Kuhns}}, \bibinfo {author} {\bibfnamefont {R.~R.}\ \bibnamefont {Urbano}},
  \bibinfo {author} {\bibfnamefont {C.}~\bibnamefont {Giles}}, \ and\ \bibinfo
  {author} {\bibfnamefont {P.~G.}\ \bibnamefont {Pagliuso}},\ }\href@noop {}
  {\bibfield  {journal} {\bibinfo  {journal} {Physical Review B}\ }\textbf
  {\bibinfo {volume} {90}},\ \bibinfo {pages} {235120} (\bibinfo {year}
  {2014})}\BibitemShut {NoStop}%
\bibitem [{\citenamefont {Thomas}\ \emph {et~al.}(2016)\citenamefont {Thomas},
  \citenamefont {Rosa}, \citenamefont {Lee}, \citenamefont {Parameswaran},
  \citenamefont {Fisk},\ and\ \citenamefont {Xia}}]{Thomas:2015tya}%
  \BibitemOpen
  \bibfield  {author} {\bibinfo {author} {\bibfnamefont {S.~M.}\ \bibnamefont
  {Thomas}}, \bibinfo {author} {\bibfnamefont {P.~F.~S.}\ \bibnamefont {Rosa}},
  \bibinfo {author} {\bibfnamefont {S.~B.}\ \bibnamefont {Lee}}, \bibinfo
  {author} {\bibfnamefont {S.~A.}\ \bibnamefont {Parameswaran}}, \bibinfo
  {author} {\bibfnamefont {Z.}~\bibnamefont {Fisk}}, \ and\ \bibinfo {author}
  {\bibfnamefont {J.}~\bibnamefont {Xia}},\ }\href {\doibase
  10.1103/PhysRevB.93.075149} {\bibfield  {journal} {\bibinfo  {journal} {Phys.
  Rev. B}\ }\textbf {\bibinfo {volume} {93}},\ \bibinfo {pages} {075149}
  (\bibinfo {year} {2016})}\BibitemShut {NoStop}%
\bibitem [{\citenamefont {Seibel}\ \emph {et~al.}(2015)\citenamefont {Seibel},
  \citenamefont {Xie}, \citenamefont {Gibson},\ and\ \citenamefont
  {Cava}}]{Seibel:2015go}%
  \BibitemOpen
  \bibfield  {author} {\bibinfo {author} {\bibfnamefont {E.~M.}\ \bibnamefont
  {Seibel}}, \bibinfo {author} {\bibfnamefont {W.}~\bibnamefont {Xie}},
  \bibinfo {author} {\bibfnamefont {Q.~D.}\ \bibnamefont {Gibson}}, \ and\
  \bibinfo {author} {\bibfnamefont {R.~J.}\ \bibnamefont {Cava}},\ }\href@noop
  {} {\bibfield  {journal} {\bibinfo  {journal} {Journal of Solid State
  Chemistry}\ }\textbf {\bibinfo {volume} {230}},\ \bibinfo {pages} {318}
  (\bibinfo {year} {2015})}\BibitemShut {NoStop}%
\bibitem [{\citenamefont {Balicas}\ \emph {et~al.}(2005)\citenamefont
  {Balicas}, \citenamefont {Nakatsuji}, \citenamefont {Lee}, \citenamefont
  {Schlottmann}, \citenamefont {Murphy},\ and\ \citenamefont
  {Fisk}}]{Balicas:2005eq}%
  \BibitemOpen
  \bibfield  {author} {\bibinfo {author} {\bibfnamefont {L.}~\bibnamefont
  {Balicas}}, \bibinfo {author} {\bibfnamefont {S.}~\bibnamefont {Nakatsuji}},
  \bibinfo {author} {\bibfnamefont {H.}~\bibnamefont {Lee}}, \bibinfo {author}
  {\bibfnamefont {P.}~\bibnamefont {Schlottmann}}, \bibinfo {author}
  {\bibfnamefont {T.}~\bibnamefont {Murphy}}, \ and\ \bibinfo {author}
  {\bibfnamefont {Z.}~\bibnamefont {Fisk}},\ }\href@noop {} {\bibfield
  {journal} {\bibinfo  {journal} {Physical Review B}\ }\textbf {\bibinfo
  {volume} {72}},\ \bibinfo {pages} {064422} (\bibinfo {year}
  {2005})}\BibitemShut {NoStop}%
\bibitem [{\citenamefont {Rodriguez}\ \emph {et~al.}(2008)\citenamefont
  {Rodriguez}, \citenamefont {Adler}, \citenamefont {Brand}, \citenamefont
  {Broholm}, \citenamefont {Cook}, \citenamefont {Brocker}, \citenamefont
  {Hammond}, \citenamefont {Huang}, \citenamefont {Hundertmark}, \citenamefont
  {Lynn}, \citenamefont {Maliszewskyj}, \citenamefont {Moyer}, \citenamefont
  {Orndorff}, \citenamefont {Pierce}, \citenamefont {Pike}, \citenamefont
  {Scharfstein}, \citenamefont {Smee},\ and\ \citenamefont
  {Vilaseca}}]{Rodriguez:2008be}%
  \BibitemOpen
  \bibfield  {author} {\bibinfo {author} {\bibfnamefont {J.~A.}\ \bibnamefont
  {Rodriguez}}, \bibinfo {author} {\bibfnamefont {D.~M.}\ \bibnamefont
  {Adler}}, \bibinfo {author} {\bibfnamefont {P.~C.}\ \bibnamefont {Brand}},
  \bibinfo {author} {\bibfnamefont {C.}~\bibnamefont {Broholm}}, \bibinfo
  {author} {\bibfnamefont {J.~C.}\ \bibnamefont {Cook}}, \bibinfo {author}
  {\bibfnamefont {C.}~\bibnamefont {Brocker}}, \bibinfo {author} {\bibfnamefont
  {R.}~\bibnamefont {Hammond}}, \bibinfo {author} {\bibfnamefont
  {Z.}~\bibnamefont {Huang}}, \bibinfo {author} {\bibfnamefont
  {P.}~\bibnamefont {Hundertmark}}, \bibinfo {author} {\bibfnamefont {J.~W.}\
  \bibnamefont {Lynn}}, \bibinfo {author} {\bibfnamefont {N.~C.}\ \bibnamefont
  {Maliszewskyj}}, \bibinfo {author} {\bibfnamefont {J.}~\bibnamefont {Moyer}},
  \bibinfo {author} {\bibfnamefont {J.}~\bibnamefont {Orndorff}}, \bibinfo
  {author} {\bibfnamefont {D.}~\bibnamefont {Pierce}}, \bibinfo {author}
  {\bibfnamefont {T.~D.}\ \bibnamefont {Pike}}, \bibinfo {author}
  {\bibfnamefont {G.}~\bibnamefont {Scharfstein}}, \bibinfo {author}
  {\bibfnamefont {S.~A.}\ \bibnamefont {Smee}}, \ and\ \bibinfo {author}
  {\bibfnamefont {R.}~\bibnamefont {Vilaseca}},\ }\href@noop {} {\bibfield
  {journal} {\bibinfo  {journal} {Measurement Science and Technology}\ }\textbf
  {\bibinfo {volume} {19}},\ \bibinfo {pages} {034023} (\bibinfo {year}
  {2008})}\BibitemShut {NoStop}%
\bibitem [{\citenamefont {Lefmann}\ \emph {et~al.}(2006)\citenamefont
  {Lefmann}, \citenamefont {Niedermayer}, \citenamefont {Abrahamsen},
  \citenamefont {Bahl}, \citenamefont {Christensen}, \citenamefont {Jacobsen},
  \citenamefont {Larsen}, \citenamefont {H{\"a}fliger}, \citenamefont
  {Filges},\ and\ \citenamefont {R{\o}nnow}}]{Lefmann:2006ek}%
  \BibitemOpen
  \bibfield  {author} {\bibinfo {author} {\bibfnamefont {K.}~\bibnamefont
  {Lefmann}}, \bibinfo {author} {\bibfnamefont {C.}~\bibnamefont
  {Niedermayer}}, \bibinfo {author} {\bibfnamefont {A.~B.}\ \bibnamefont
  {Abrahamsen}}, \bibinfo {author} {\bibfnamefont {C.~R.~H.}\ \bibnamefont
  {Bahl}}, \bibinfo {author} {\bibfnamefont {N.~B.}\ \bibnamefont
  {Christensen}}, \bibinfo {author} {\bibfnamefont {H.~S.}\ \bibnamefont
  {Jacobsen}}, \bibinfo {author} {\bibfnamefont {T.~L.}\ \bibnamefont
  {Larsen}}, \bibinfo {author} {\bibfnamefont {P.}~\bibnamefont
  {H{\"a}fliger}}, \bibinfo {author} {\bibfnamefont {U.}~\bibnamefont
  {Filges}}, \ and\ \bibinfo {author} {\bibfnamefont {H.~M.}\ \bibnamefont
  {R{\o}nnow}},\ }\href@noop {} {\bibfield  {journal} {\bibinfo  {journal}
  {Physica B: Condensed Matter}\ }\textbf {\bibinfo {volume} {385-386}},\
  \bibinfo {pages} {1083} (\bibinfo {year} {2006})}\BibitemShut {NoStop}%
\bibitem [{\citenamefont {Pelissetto}\ and\ \citenamefont
  {Vicari}(2002)}]{Pelissetto2002549}%
  \BibitemOpen
  \bibfield  {author} {\bibinfo {author} {\bibfnamefont {A.}~\bibnamefont
  {Pelissetto}}\ and\ \bibinfo {author} {\bibfnamefont {E.}~\bibnamefont
  {Vicari}},\ }\href {\doibase http://dx.doi.org/10.1016/S0370-1573(02)00219-3}
  {\bibfield  {journal} {\bibinfo  {journal} {Physics Reports}\ }\textbf
  {\bibinfo {volume} {368}},\ \bibinfo {pages} {549 } (\bibinfo {year}
  {2002})}\BibitemShut {NoStop}%
\bibitem [{\citenamefont {Le~Guillou}\ and\ \citenamefont
  {Zinn-Justin}(1980)}]{PhysRevB.21.3976}%
  \BibitemOpen
  \bibfield  {author} {\bibinfo {author} {\bibfnamefont {J.~C.}\ \bibnamefont
  {Le~Guillou}}\ and\ \bibinfo {author} {\bibfnamefont {J.}~\bibnamefont
  {Zinn-Justin}},\ }\href {\doibase 10.1103/PhysRevB.21.3976} {\bibfield
  {journal} {\bibinfo  {journal} {Phys. Rev. B}\ }\textbf {\bibinfo {volume}
  {21}},\ \bibinfo {pages} {3976} (\bibinfo {year} {1980})}\BibitemShut
  {NoStop}%
\bibitem [{\citenamefont {Chattopadhyay}\ \emph {et~al.}(1994)\citenamefont
  {Chattopadhyay}, \citenamefont {Burlet}, \citenamefont {Rossat-Mignod},
  \citenamefont {Bartholin}, \citenamefont {Vettier},\ and\ \citenamefont
  {Vogt}}]{PhysRevB.49.15096}%
  \BibitemOpen
  \bibfield  {author} {\bibinfo {author} {\bibfnamefont {T.}~\bibnamefont
  {Chattopadhyay}}, \bibinfo {author} {\bibfnamefont {P.}~\bibnamefont
  {Burlet}}, \bibinfo {author} {\bibfnamefont {J.}~\bibnamefont
  {Rossat-Mignod}}, \bibinfo {author} {\bibfnamefont {H.}~\bibnamefont
  {Bartholin}}, \bibinfo {author} {\bibfnamefont {C.}~\bibnamefont {Vettier}},
  \ and\ \bibinfo {author} {\bibfnamefont {O.}~\bibnamefont {Vogt}},\ }\href
  {\doibase 10.1103/PhysRevB.49.15096} {\bibfield  {journal} {\bibinfo
  {journal} {Phys. Rev. B}\ }\textbf {\bibinfo {volume} {49}},\ \bibinfo
  {pages} {15096} (\bibinfo {year} {1994})}\BibitemShut {NoStop}%
\bibitem [{\citenamefont {Seabra}\ \emph {et~al.}(2016)\citenamefont {Seabra},
  \citenamefont {Sindzingre}, \citenamefont {Momoi},\ and\ \citenamefont
  {Shannon}}]{Seabra:2016ba}%
  \BibitemOpen
  \bibfield  {author} {\bibinfo {author} {\bibfnamefont {L.}~\bibnamefont
  {Seabra}}, \bibinfo {author} {\bibfnamefont {P.}~\bibnamefont {Sindzingre}},
  \bibinfo {author} {\bibfnamefont {T.}~\bibnamefont {Momoi}}, \ and\ \bibinfo
  {author} {\bibfnamefont {N.}~\bibnamefont {Shannon}},\ }\href@noop {}
  {\bibfield  {journal} {\bibinfo  {journal} {Physical Review B}\ }\textbf
  {\bibinfo {volume} {93}},\ \bibinfo {pages} {085132} (\bibinfo {year}
  {2016})}\BibitemShut {NoStop}%
\end{thebibliography}%
\end{document}